\RequirePackage{fix-cm}
\documentclass[smallextended]{svjour3}       
\pdfoutput=1
\smartqed  
\usepackage{color}

\usepackage{graphicx}

\newcommand{\spin}{\sigma}
\newcommand{\ospin}{\overline{\sigma}}
\newcommand{\kv}{{\bf k}}
\newcommand{\qv}{{\bf q}}

\begin{document}

\title{Superconducting phase diagram of the paramagnetic one-band Hubbard model
\thanks{B.M.A., A.T.R. and A.K. acknowledge support from a Lundbeckfond fellowship (Grant A9318).  P.J.H. was supported by NSF-DMR-1407502.
}}


\author{Andreas Kreisel$^{1}\!$ \and 
        Astrid T. R\o mer$^{1}\!$ \and 
        P. J. Hirschfeld$^2\!$ \and 
        Brian M. Andersen$^{1}$
}

\authorrunning{A. Kreisel, A.T. R\o mer, P.J. Hirschfeld, B.M. Andersen} 

\institute{$^1$Niels Bohr Institute, University of Copenhagen, DK-2100 Copenhagen, Denmark\\
$^2$ Department of Physics, University of Florida, Gainesville, Florida 32611, USA
}

\date{September 6, 2016}

\maketitle

\begin{abstract}
We study spin-fluctuation-mediated superconductivity in the one-band Hubbard model. Higher order effective interactions in $U$ give rise to a superconducting instability which is very sensitive to changes in the Fermi surface topology arising as a function of doping and changes in the band structure. We show the superconducting phase diagram as a function of doping and next-nearest neighbor hopping in the limit of very small Coulomb interaction strength and discuss peculiarities arising at the phase boundaries separating different superconducting domains. \\
\keywords{Hubbard model, Spin-fluctuation-mediated superconductivity}
\end{abstract}

\section{Introduction}
\label{intro}
The physics of the repulsive Hubbard model includes the Mott phase, magnetism, stripe spin- and charge-ordered phases, and superconductivity\cite{dagotto,bianconi,kivelson,leewen}. The weak-coupling approach to superconductivity within the one-band Hubbard model highlights the role of spin fluctuations\cite{berkschrie}. It is well-known that close to half filling, $\langle n \rangle =1$, for  next-nearest hopping constants $|t^\prime|<0.5t$, proximity to an antiferromagnetic instability promotes $d_{x^2-y^2}$ superconductivity \cite{Scalapino86}, but other regions of the phase space $(\langle n\rangle, |t^\prime|)$ have been less extensively studied. Early on it was realized that the Fermi surface topology plays a most important role for the pairing symmetry \cite{Scalapino87} due to the different nesting conditions arising when band structure and filling are varied. Thus, the one-band Hubbard model hosts many different possible superconducting instabilities. A comprehensive study was carried out by Hlubina \cite{Hlubina99}, who found that triplet superconductivity is in fact dominant in a large region near $\langle n \rangle =0.5$ and for $|t^\prime|<0.5t$; results for larger $|t'|$ can be found for example in Ref. \cite{Hassan08}. This promotion of triplet superconductivity is correlated with positions in phase space where a van Hove singularity resides at or very close to the Fermi level~\cite{Astrid15,Raghu} and thus occurs at the verge of a ferromagnetic instability.
In a recent study\cite{Astrid15}, we calculated the pairing symmetry of the one-band Hubbard model in the paramagnetic limit for a large range of fillings $\langle n \rangle =0-1.5$ and next-nearest neighbor hopping constants, $|t^\prime|=0-1.5$. The leading superconducting instability was found for intermediate Coulomb interaction strength, $U \simeq 1-3$ ensuring a non-negligible critical temperature. It is possible, however,  that changes in the Coulomb strength can affect the balance between near-lying instabilities and even interchange the leading and sub-leading instabilities. Therefore, results from the pertubative limit, i.e. $U\to 0$\cite{Hlubina99,Simkovic} might differ from the results for intermediate coupling strengths. In this paper, we show that the results of Ref.~\cite{Astrid15} with intermediate interaction strengths are in overall agreement with the $U\to 0$ limit, despite some small discrepancies in boundary regions which separate domains of different pairing symmetry. We discuss these variations and show examples of small differences in the boundary regions of the phase diagram as a function of $U$. Furthermore, we show that the boundary regions are particularly interesting, since they provide the possibility of time-reversal-symmetry-broken (TRSB) superconducting phases which can be realized as a consequence of two nearly degenerate solutions.

\section{Model and Method}
\label{sec:model}
We consider the Hubbard model for a two-dimensional square lattice
\begin{equation}
 H=\sum_{\kv \spin}\xi_kc_{\kv\spin}^\dagger c_{\kv\spin}+\frac{U}{2N}\sum_{\kv,\kv',\qv}\sum_{\spin}c_{\kv'\spin}^\dagger c_{-\kv'+\qv\ospin}^\dagger c_{-\kv+\qv\ospin} c_{\kv\spin},
\end{equation}
where $\xi_\kv=-2t[\cos(k_x)+\cos(k_y)]-4t^\prime\cos(k_x)\cos(k_y)-\mu$ with $t$ being the hopping integral to nearest neighbors, and $t^\prime<0$ the hopping integral between next-nearest neighbors. In the following we set $t=1$ and restrict ourselves to the case of negative values of $t'$ with $|t^\prime|=0-1.5$.

A spin-fluctuation-mediated interaction can combine two electrons of opposite or same spin into a Cooper pair.
Higher order diagrams of the repulsive Coulomb interaction $U$ are used to derive the pairing interaction~\cite{berkschrie,Scalapino86},
\begin{eqnarray}
 \Gamma_{\kv,\kv'}^\textrm{\small opposite spin}&=&U+\frac{U^2}{2}\chi_{(\kv-\kv')}^{sp}-\frac{U^2}{2}\chi_{(\kv-\kv')}^{ch}+U^2\chi_{(\kv+\kv')}^{sp},  \label{eq:PMoppspin}  \\
 \Gamma_{\kv,\kv'}^\textrm{\small same spin}&=&-\frac{U^2}{2}\chi_{(\kv-\kv')}^{sp}-\frac{U^2}{2}\chi_{(\kv-\kv')}^{ch},  \label{eq:PMsamespin}
 \end{eqnarray}
 with the spin susceptibility $ \chi^{sp}_\qv=\chi_0(\qv)[1-U\chi_0(\qv)]^{-1}$ and the charge susceptibility $\chi^{ch}_\qv=\chi_0(\qv)[1+U\chi_0(\qv)]^{-1}$.
Equations~(\ref{eq:PMoppspin}) and~(\ref{eq:PMsamespin}) provide a measure of the interaction strength, and we neglect the energy dependence of the interactions.
Evaluating the Lindhard function
$\chi_0(\qv,\omega)=1/N\sum_\kv [f(\xi_{\kv+\qv})-f(\xi_{\kv})][\omega+\xi_\kv-\xi_{\kv+\qv}+i\eta]^{-1}$
at zero energy ($\omega=0$), we obtain the bare susceptibility in the paramagnetic phase.
 The gap equation is determined by calculation of the effective pair scattering vertex  in the random phase approximation (RPA).  In the singlet $(s)$ and triplet $(t)$ channel it takes the form
\begin{eqnarray}
  \Delta^{s/t}_\kv&=&
  -\frac{1}{2N}\sum_{\kv'}\Gamma^{s/t}_{\kv,\kv'}\frac{\Delta^{s/t}_{\kv'}}{2E^{s/t}_{\kv'}}\tanh\Big(\frac{E^{s/t}_{\kv'}}{2k_BT}\Big),
\label{eq:SCGapEquationNS}
\end{eqnarray}
with
$E_\kv^{s/t}=\sqrt{\xi_{\kv}^2+|\Delta^{s/t}_\kv|^2}$.
In the calculation of the superconducting gap, the potential forms stated in Eqs.~(\ref{eq:PMoppspin}) and ~(\ref{eq:PMsamespin}) must be symmetrized or antisymmetrized with respect to momentum in the even-parity singlet and odd-parity triplet channel, respectively. In the case of opposite spin electrons we thus have
\begin{equation}
\Gamma^{s/t}_{\kv,\kv'}=\Gamma^\textrm{\small opposite spin}_{\kv,\kv'}\pm \Gamma^\textrm{\small opposite spin}_{-\kv,\kv'}.
\end{equation}
Note that the potential entering Eq.~(\ref{eq:SCGapEquationNS}) appears in the singlet (even in $\kv$) and triplet (odd in $\kv$) form explicitly. This symmetry directly carries over to the gap, ensuring that $\Delta_{\kv}^s=\Delta_{-\kv}^s$ and $\Delta_\kv^t=-\Delta_{-\kv}^t$.

\begin{figure*}
  \includegraphics[width=\linewidth]{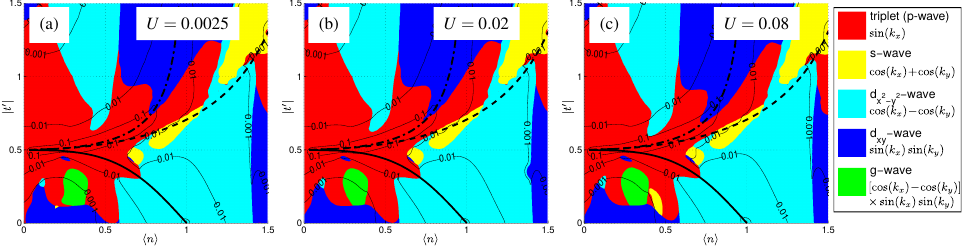}
\caption{Phase diagram in the small $U$ limit for three different values of the interaction showing small differences of the phase boundaries, e.g. changes of the order of the superconducting instabilities.
For direct comparison to the small $U$ limit as presented in \cite{Simkovic}, the \textit{contour lines} show the value of the eigenvalue divided by the interaction $\tilde\lambda=\lambda/U^2$.}
\label{fig:1}       
\end{figure*}

In the limit $T\rightarrow T_c$, $\Delta_{\kv}\rightarrow 0$, we solve the linearized gap equation in the singlet and triplet channels
\begin{equation}
\Big[ -\frac{1}{2(2\pi)^2}\int_{FS}\frac{d\kv'}{|v_{\kv'}|}\Gamma^{s/t}_{\kv,\kv'}\Big]g(\kv')= \lambda g(\kv),
\label{eq:lge}
\end{equation}
by diagonalization of the matrix
\begin{equation}
M_{\kv,\kv'}= -\frac{1}{2(2\pi)^2}\frac{l_{\kv'}}{|v_{\kv'}|}\Gamma^{s/t}_{\kv,\kv'}.
\end{equation}
Here $\kv$ and $\kv'$ are located on the Fermi surface and $l_\kv$ is the length of the Fermi surface segment associated with the point $\kv$ while $v_{\kv}$ is the Fermi velocity. By this procedure we 
identify the leading instability with gap symmetry function $g(\kv)$ by the largest eigenvalue $\lambda$ as a function of electron filling and next-nearest neighbor hopping constant, $t'$.
The leading eigenfunction is characterized according to its transformation properties and labeled by one of the irreducible representations of the $D_{4h}$ group that are even under reflection
through the horizontal plane, $s,d_{x^2-y^2},d_{xy},g,p$, see Fig. \ref{fig:1}.

\section{Results}
\label{sec:results}
\begin{figure*}
  \includegraphics[width=\linewidth]{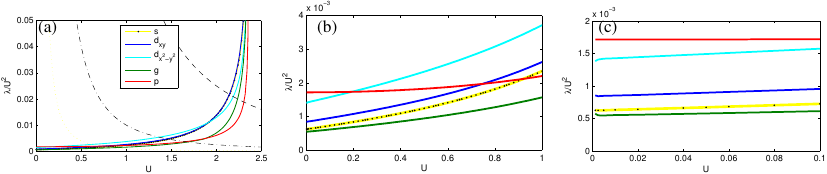}
\caption{Pairing strength $\lambda/U^2$ in the different symmetry channels as a function of the interaction $U$ for $\langle n \rangle =0.55$ and $|t^\prime|=0.25$. (a) For large interactions, the divergence of the pairing eigenvalue $\lambda$ shows the breakdown of the
weak-coupling approach. Sizable pairing eigenvalues of $\lambda=0.001$ and $\lambda=0.01$ are reached at the intersection with the dash-dotted and dashed lines. (b) For interaction strengths $U<1$, we observe a dependence of the leading instability on the value of $U$. This can lead to moving phase boundaries.  (c) In the limit $U\rightarrow 0$ the order of the instabilities
does not change. However, the pairing eigenvalue is very small corresponding to an exponentially small critical temperature $T_c$.}
\label{fig_lambda_u}       
\end{figure*}

\paragraph{Phase diagram in the small $U$ limit.}
In Fig. \ref{fig:1} we show the superconducting phase diagram for fillings in the range $\langle n \rangle=0-1.5$ and next-nearest hopping constants $|t^\prime|=0-1.5$ in the limit of very small Coulomb interaction, $U\rightarrow 0$. As seen from Fig. \ref{fig:1}, the region of triplet superconductivity shrinks upon increasing the Coulomb interaction, but the main features of the phase diagram are robust to changes in the Coulomb strength. 
In a previous work \cite{Astrid15}, we studied the phase diagram for larger Coulomb interactions corresponding to sizable value of the superconducting critical temperature. The phase diagrams of either approach display an overall agreement despite a substantial reduction of the triplet superconductivity area with increasing $U$, as well as some new superconducting phases which appear in the boundary regions separating different superconducting domains.
This shows that increasing the Coulomb interaction can lead to a change in symmetry of the leading superconducting instability. This often happens close to a magnetic instability, where the peaks in the susceptibility are very sharp and can boost one or the other superconducting state
depending on the available states on the Fermi surface and their momenta $\bf k$ and $\bf k^\prime$, see Fig. \ref{fig_lambda_u} (a). The weak-coupling approach breaks down for large interactions and correlations eventually remove these singularities; however an interchange between the leading instabilities can occur at rather small interactions as shown for a suitable region in the phase diagram (Fig. \ref{fig_lambda_u} (b)). Therefore, the realistic situation with finite $T_c$ might appear different from the estimate in the small $U$ limit
where perturbation theory is valid (Fig. \ref{fig_lambda_u} (c)). An example of this is the shrinkage of the triplet phase upon increased interaction strength, which happens due to the promotion of the peak structure of the susceptibility making the singlet solutions more competitive. Moving towards a more realistic scenario with sizable eigenvalues $\lambda=\mathcal{O}(1)$, new phases (for example, the phase of s-wave symmetry close to filling $\langle n\rangle=0.5$ and $|t'|=0.25$) show up. The case for $t'=0$ is of course particle-hole symmetric for reflections at $\langle n\rangle =1$ and agrees in the small $U$ limit (Fig. \ref{fig_lambda_u} (a)) surprisingly well with the sequence of phases found in a recent Monte Carlo investigation\cite{Deng15}. The exception is the limit of extremely small  densities our method is less reliable and shows deviations to limits worked out analytically previously \cite{Deng15,Chubukov92}.
\begin{figure*}
\includegraphics[width=\linewidth]{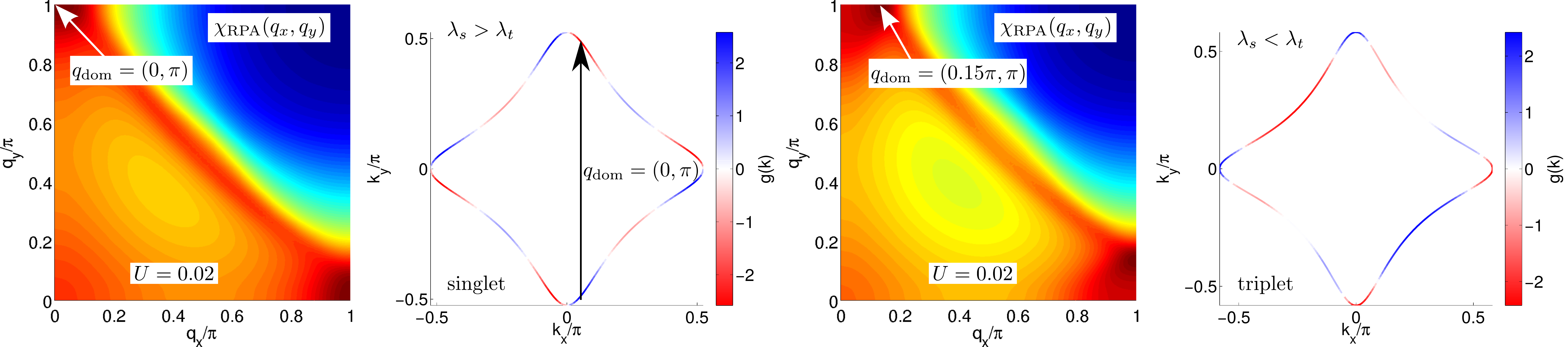}
\caption{The superconducting instability of $d_{xy}$ symmetry near $|t'|=0.44$ and $\langle n \rangle=0.26$ in the phase diagram is stable because of the commensurate structure of the susceptibility.
The dominating peak in the susceptibility is at $q_{\mathrm{dom}}=(0,\pi)$ (and symmetry
related vectors) such that Cooper pairs at some parts of the Fermi surface can take advantage of the corresponding pair-scattering amplitude leading to a larger singlet instability $\lambda_s>\lambda_d$ (left). Moving towards a larger filling $n=0.3$ also moves the
dominating peak to an incommensurate value such that $\lambda_s<\lambda_d$ and the instability towards triplet superconductivity is stronger (right).}
\label{fig_d_o}       
\end{figure*}

\paragraph{Robust phase with commensurate structure of the susceptibility.}
Interestingly, a small phase of $d_{xy}$ symmetry is present at all values of the interaction \cite{Astrid15,Simkovic} fixed at the position $\langle n\rangle\approx0.26$ and $|t^\prime|\approx 0.44$. This small region is surrounded by triplet superconductivity and therefore appears to be a special point in the phase diagram. Examining the susceptibility at that point, one sees that it has a peak at the commensurate vector $q_{\mathrm{dom}}=(0,\pi)$. When changing filling or next-nearest neighbor hopping $t^\prime$ the peak moves to an incommensurate position $(a,\pi)$ or $(0,\pi-b)$ (see Fig. \ref{fig_d_o}) and thereby renders the singlet instability less favorable.

\paragraph{Time-reversal-symmetry-broken state.}
\begin{figure*}
\includegraphics[width=\linewidth]{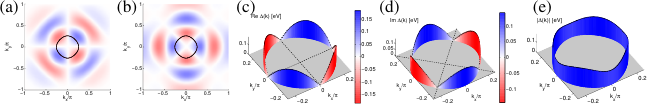}
\caption{Solution of the full BdG equation at $k_BT=0.015$ for low filling $\langle n\rangle =0.1$, $|t'|=0.25$ and Coulomb interaction strength $U=5.23$ yielding a leading pairing eigenvalue of $\lambda=0.2$ with dominant $d_{x^2-y^2}$ eigenfunction and sub-dominant $d_{xy}$ symmetry: The self-consistent solution is a TRSB state with the real part of the gap in the whole Brillouin zone of $d_{xy}$ symmetry (a) and the imaginary part (b) of $d_{x^2-y^2}$ symmetry; the black line is the Fermi surface. Furthermore, we show the real part of the gap at the Fermi surface (c) and the imaginary part (d) as well as the finite spectral gap $|\Delta(k)|$ (e).}
\label{fig_trsb}       
\end{figure*}
Generally, if the pairing interaction allows for two superconducting instabilities which are degenerate or nearly degenerate, the possibility of time-reversal-symmetry-broken (TRSB) states arises. Such a degeneracy naturally occurs in the case of triplet pairing because pairs of eigenfunctions are degenerate by symmetry. A complex superposition of these states, e.g. $p_x+ip_y$, remove nodes from the Fermi surface and has a lower energy than either $p_x$ or $p_y$ alone\cite{Astrid15}. In the singlet channel, each symmetry channel is not degenerate. Therefore, a TRSB ground state is in general not realized, but may develop at low temperatures in regions of the phase diagram where near-degeneracy between two different singlet solutions of the linearized gap equation occur, i.e. close to a phase boundary. In this case, the removal of nodes at the Fermi surface causes a gain in condensation energy. For our present one-band model, we find a near-degeneracy between $d_{x^2-y^2}$ and $d_{xy}$ symmetry at small fillings and sizable Hubbard interaction. A self-consistent Bogoliubov-de Gennes (BdG) calculation selects a $d_{xy}+id_{x^2-y^2}$ order parameter as shown in Fig. \ref{fig_trsb} (a-d). The nodal lines of the two symmetries (see Fig. \ref{fig_trsb} (c-d)) cross the Fermi surface at different positions and the complex superposition therefore leads to a finite gap at all points of the Fermi surface. This gives rise to the full gap as shown in Fig. \ref{fig_trsb} (e).

\section{Conclusions}
We have shown how changes in Fermi surface topology govern the phase diagram of the superconducting instability in the weak-coupling approach to the Hubbard model. In addition, we have also provided a detailed discussion of how changes in the Coulomb interaction strength may affect the superconducting pairing problem and shift the balance between leading and sub-leading instabilities of the linearized gap equation. However, robust features of the superconducting phase diagrams are present when the susceptibility displays commensurate features, as was shown in the case of the $d_{xy}$ superconducting island inside a domain of triplet superconductivity.

Phase boundary regions separating different superconducting domains are special in the sense that two or more superconducting instabilities are nearly degenerate. This can pave the way for more exotic superconducting phases. Specifically, we showed that such a boundary region hosts the TRSB superconducting gap $d_{xy}+id_{x^2-y^2}$. 



\end{document}